\documentstyle[aps]{revtex}

\newcommand{\be}{\begin{equation}}
\newcommand{\ee}{\end{equation}}
\newcommand{\beq}{\begin{eqnarray}}
\newcommand{\eeq}{\end{eqnarray}}

\newcommand{\bF}{{\cal F}}
\newcommand{\ffi}{\varphi}
\newcommand{\eps}{\varepsilon}
\newcommand{\ps}{\psi'}

\begin{document}
%\
%\vspace{2cm}
%\begin{center} \Large \bf
\title{ The effective action and quantum gauge transformations}
%\end{center}
%\centerline{S.Alexandrov}
\author{S.~Alexandrov\thanks{e.mail: alexand@snoopy.phys.spbu.ru}}

\address{Department of Theoretical Physics,
St.Petersburg University,198904 St.Petersburg, Russia}

%\begin{center}{\it Department of Theoretical Physics,
%St.Petersburg University,}\\{\it 198904 St.Petersburg, Russia}
%\end{center}

\maketitle

\begin{abstract}
 The local symmetry transformations of the quantum effective action
for general gauge theory are found. Additional symmetries arise
under consideration of background gauges. Together with
"trivial" gauge transformations, vanishing on mass shell,
they can be used for construction simple gauge generators.
For example, for the Yang--Mills theory the classically invariant
effective action is obtained, reproducing DeWitt's result.
For rank one theories a natural generalization is proposed.

\end{abstract}

\section{Introduction}

  The concept of symmetry was and remains a very powerful tool
for construction of the quantum field theory. One of its main virtues
is that the symmetry restricts a form of the action, which lies
in the ground of the theory. Consequences of the classical
symmetry play a crucial role  for renormalizability of the quantum
theory. And in the investigation of this problem the effective action
takes a very important place \cite{renom}. Besides it is the only
quantum object in which the symmetry should be reflected by the same
way as in the classical action also restricting the number of available
structures. So it is natural to find this quantum realization
of the symmetry, i.e. the symmetry transformations of the effective
action, in an explicit form.

 One of first steps in this direction was done by DeWitt in his
construction of the classically gauge invariant effective action
for  the Yang--Mills theory \cite{dewitt}. This work gave rise to
the number of papers devoted to this problem \cite{all}.
But all of them do not go beyond linear gauge transformations and
background gauges of the certain kind. This is a very strong limitation
on the physical theory. As we know the Hamiltonian forms of gravity,
supersymmetry theories and many others require nonlinear
transformations. So an investigation of general gauge theories
from the point of view of the quantum gauge symmetry is needed.

 The following break-through is connected with the concept of
the effective average action or Vilkovisky--DeWitt's action
\cite{vilk2}--\cite{dewi}. Its gauge invariance and gauge independence
are very attractive properties.
However its actual construction in arbitrary
gauge and for arbitrary gauge theory is an enormously hard task
because the connection on the frame bundle on the space--of--histories
is needed. Besides
the effective average action is connected to the ordinary generating
functional for the one--particle--irreducible Green functions
in a nontrivial way.
So we simply bypass the subject and
consider the effective action constructed in the usual way as
Legendre transformation.

 The common approach to the symmetry properties of the effective
action for general gauge theories is investigation of
the Ward identities (see for example \cite{lavr},\cite{lv}).
They are the reflection of the global BRST symmetry
which replaces the gauge symmetry in path integral quantization
\cite{BRST}. This symmetry plays a leading role in quantization of
general gauge theories being the basis for Hamiltonian BFV \cite{BFV}
and Lagrangian BV \cite{bv} quantization schemes. Within these
approaches global symmetry transformations of the effective action,
which are called quantum BRST transformations, can easily be found
\cite{vlt,qtrans,bigrev,htbook}.
Here we are interested in their local counterparts which
are realized on the physical fields only.
Explicit formulas for them, to our knowledge,
are absent in the literature and
our aim is to fill in this gap.
Besides we discuss the symmetry transformations in presence of
background fields and apply the obtained results to
the rank one theory.

 Our consideration is based on the Hamiltonian form of
BRST quantization.  An alternative strategy is to follow
the BV formalism. Some remarks on this point are given in Sec.VI.

 The paper is organized as follows. In Sec.II the BFV quantization
is reviewed and for completeness and to fix the notations
the quantum BRST transformations are obtained.
In the subsequent section
the gauge transformations of the effective action in terms of
quantum averages are found and the "trivial" transformations and
the symmetry algebra are discussed. In Sec.IV it is shown that
introduction of background fields results in appearance of additional
symmetries which can be combined with initial ones. This is used in
Sec.V to construct the symmetry, which is reduced to the classical
gauge transformations in the case of the  Yang--Mills theory
reproducing DeWitt's result and gives a generalization for
nonlinear transformations. In the last section some problems
and perspectives are outlined.

 Our condensed notations correspond to DeWitt's ones introduced in
\cite{book}. This may lead to confusion when they are applied,
for example, in an expression for the classical action. In such cases
all tensors must be understood as local what results in locality of
the whole expression. In this connection it is convenient to extend
the definition of the Poisson brackets on variables depending on
different moments of time. We demand  $\{q^s,p_r\}=
\delta^s_r$, i.e. an unsimultaneous commutator vanishes. This
provides locality and keeps unity of the notations. All derivatives
with respect to Grassmann variables are left and
for simplicity we restrict ourselves to the case of even classical
fields only.

\section{Preliminaries}

 Let us survey the Hamiltonian BFV quantization following
the review \cite{Henneaux}.

Consider a gauge theory with  phase space
variables $z^A=(q^s,p_s)$,
Hamiltonian $H_0(q,p)$ and first class constraints $G_{\alpha}$.
Let $n^{\alpha}$ be the Lagrange multipliers associated with
the constraints $G_{\alpha}$, and $\pi_{\alpha}$ be the canonically
conjugate momenta. The action is given by
\be S(q,p,n)=\dot q^sp_s-H_0-n^{\alpha}G_{\alpha}
\label{Sg}, \ee
whereas the gauge transformations are
$ \delta \ffi^a=G^a_{\alpha}(\ffi)\eps^{\alpha}$,
where $\ffi^a=(q^s,p_s,n^{\alpha})$ and
\be G^{(z^A)}_{\alpha}=\{ z^A,G_{\alpha} \},\qquad
  G^{(n^{\beta})}_{\alpha}=\delta^{\beta}_{\alpha}\partial_t+n^{\gamma}
 C^{\beta}_{\alpha \gamma}-V^{\beta}_{\alpha} \label{gtr} \ee
with $C^{\gamma}_{\alpha\beta},V_{\alpha}^{\beta}$ defined through
the relations
$\{ G_\alpha ,G_\beta \} =C^{\gamma}_{\alpha\beta}G_\gamma$ and
$\{ H_0,G_{\alpha}\}=V_{\alpha}^{\beta}G_{\beta}$.
The extended phase space is defined by introducing
extra ghost and antighost fields
$(b^{\alpha},\bar c_{\alpha},c^{\alpha},\bar b_{\alpha})$.
obeying the following nonvanishing
antibrackets
$$ \{ b^{\alpha} ,\bar c_{\beta} \}_+=-\delta^{\alpha}_{\beta},\
\{ c^{\alpha} ,\bar b_{\beta} \}_+=-\delta^{\alpha}_{\beta}. $$
 $c^{\alpha},\bar c_{\alpha}$ are real,
whereas $b^{\alpha},\bar b_{\alpha}$ are imaginary.
It is convenient to define an additional structure on the extended
phase space, that of "ghost number". This is done by attributing
the following ghost number to the canonical variables:
 $c^{\alpha},b^{\alpha}$ have ghost number one,
 $\bar c_{\alpha},\bar b_{\alpha}$ have ghost number minus one.
All other variables have ghost number zero.

On this space one can construct a BRST generator $\Omega$
and a BRST invariant Hamiltonian $H$.
They are determined by the following conditions:

(a) $\Omega$ is real and odd; (b) $\Omega$ has ghost number one;
(c) $\Omega =-ib^{\alpha}\pi_{\alpha}+c^{\alpha}G_{\alpha}+
"higher\  ghost\ terms"$; (d) $\{ \Omega ,\Omega \}_+=0$

(a) $H$ is real and even; (b) $H$ has ghost number zero;
(c) $H$ coincides with $H_0$ up to higher ghost terms;
(d) $\{ H ,\Omega \}=0$

 The BRST generator is fully defined by structure functions of
the constraint algebra:
\be \Omega=-ib^{\alpha}\pi_{\alpha}+\sum\limits_{n\ge 0}c^{\alpha_{n+1}}
\cdots c^{\alpha_1}U^{(n)\beta_1 \cdots \beta_n}_{\alpha_1
\cdots \alpha_{n+1}}\bar b_{\beta_n}\cdots \bar b_{\beta_1}.
\label{Om} \ee
For $n=0$ and $n=1$ they are
$ U^{(0)}_\alpha =G_\alpha ,\
U^{(1)\gamma}_{\alpha\beta}=
-\frac 12 C^{\gamma}_{\alpha\beta}.$
Higher order structure functions are defined through repeated
Poisson brackets of the constraints. The Hamiltonian $H$ in the first
orders in ghosts has the form
 \be H=H_0+c^{\alpha}V_{\alpha}^{\beta}\bar b_{\beta}+\cdots.
 \label{Ham}  \ee

The quantization is based on the generating functional for the Green
functions which is represented in the form
\be Z[J]=e^{\frac{i}{\hbar}W(J)} =\int {\cal D}\ps\, e^{\frac{i}{\hbar}
(S_{eff}(\ps)+J_i\ps^i)}, \label{zed} \ee
where
\be  S_{eff}=\dot q^s p_s +\dot n^{\alpha}\pi_{\alpha}+
\dot c^{\alpha}\bar b_{\alpha}+\dot b^{\alpha}\bar c_{\alpha}-H_{eff},
\qquad
H_{eff}=H-\{ \Psi ,\Omega \}_+  . \label{Leff} \ee
Here $\Psi$ is an odd and imaginary function which has
ghost number minus one and plays a role of gauge fixing function,
$\psi=(\ffi,\pi,\eta=(c,b),\bar \eta=(\bar c,\bar b))$.
$\ffi^a$ are just the physical fields of the theory.
 From properties of $H$ and $\Omega$ the invariance under
the global BRST transformations follows:
$\{ S_{eff},\Omega \}=-\int dt\frac{d}{dt}\Omega =0$.

 We choose
\begin{equation}
\Psi= \bar b_{\alpha}n^{\alpha}+i\bar c_{\alpha}\chi^{\alpha}.
\label{psi}
\end{equation}
Substituting (\ref{Om}) and (\ref{psi}) in (\ref{Leff}) one obtains
\cite{htbook}:
\beq  S_{eff}&=& \dot q^s p_s +\dot n^{\alpha}\pi_{\alpha}+
\dot c^{\alpha}\bar b_{\alpha}+\dot b^{\alpha}\bar c_{\alpha}-H-
\sum\limits_{n\ge 0}(n+1)n^{\alpha_{n+1}}c^{\alpha_n}\cdots
c^{\alpha_1} U^{(n)\beta_1 \cdots \beta_n}_{\alpha_1 \cdots
\alpha_{n+1}} \bar b_{\beta_n} \cdots \bar b_{\beta_1}
\nonumber \\
 & &  -\pi_{\alpha}\chi^{\alpha}
 +ib^{\alpha}\bar b_{\alpha} -b^{\alpha}\frac{\delta\chi^{\beta}}
{\delta n^{\alpha}}\bar c_{\beta} +i\sum\limits_{n\ge 0}c^{\alpha_{n+1}}
\cdots c^{\alpha_1} \{ U^{(n)\beta_1 \cdots \beta_n}_{\alpha_1 \cdots
\alpha_{n+1}},\chi^{\gamma} \} \bar b_{\beta_n}\cdots \bar b_{\beta_1}
\bar c_{\gamma}  . \label{Seff} \eeq
This action gives rise to the BRST extended effective action
\be \Gamma_{eff}(\psi)=W(J)-J_i\psi^i,
\ \ \ \frac{\delta W(J)}{\delta J_i} =\psi^i(J). \label{ea} \ee

Let us obtain its symmetry transformations. (Following results
can be found for instance in \cite{htbook}.)
The condition that some transformation is a symmetry of the
effective action can be written as
\be \frac{\delta \Gamma_{eff}(\psi)}{\delta \psi^i}\delta \psi^i=
-J_i\delta\psi^i=0. \ee
Thus the Ward identities homogeneous
on the sources are needed. They are immediately obtained from
the path integral (\ref{zed}). Due to the BRST invariance
of the action $S_{eff}$ the change of variables
$\ps^i \to \ps^i+\Omega^i\eps$,
where $\Omega^i(\psi)=\{ \psi^i,\Omega \}$,
in the first order in $\eps$ leads to  the equality
\be J_i\Omega^i(\psi)\vert_{\psi\to \frac{\hbar\delta}
{i\delta J}} Z(J)=0 \Leftrightarrow J_i\Lambda^i=0, \label{Ward}\ee
where
\be \Lambda^i(\psi)=e^{-\frac{i}{\hbar}W(J)}\Omega^i(\psi)
\vert_{\psi\to \frac{\hbar\delta}{i\delta J}}e^{\frac{i}{\hbar}W(J)}.
\label{Lmb} \ee
 The functional $\Lambda^{i}(\psi)$ is just the quantum BRST
transformation of the effective action. This result means that the
quantum transformation is a vacuum average of T--product of
the classical one in the external field, i.e.
\be \delta^{(q)}\psi^i=\langle T\{ \delta \psi^i
\vert_{\psi=\hat \psi} \} \rangle_{J(\psi)}. \ee
(This result holds also in the BV formalism \cite{bigrev,anselm}.)
 In this connection it is convenient to introduce the following
notation $\langle f(\psi) \rangle=\langle T\{ f(\hat \psi)\}
\rangle_{J(\psi)}$.
Then one can write the simple equality
\be \Lambda^i(\psi)=\langle \Omega^i(\psi) \rangle. \label{glsym}\ee
 It is useful to emphasize that quantum corrections to the classical
transformation arise from second and higher derivatives of $W(J)$.

\section{The quantum gauge transformations}

 In the previous section we have obtained the global quantum symmetry
of the effective action. To find corresponding local transformations
it is necessary to return to the physical fields at quantum level.

   It is easy to see (see for example \cite{htbook})
that with the choice (\ref{psi}) the
 following conditions reduce (\ref{Seff}) to (\ref{Sg})
  \be \eta=\bar \eta=0, \qquad \pi=0 .  \label{cond} \ee
  The gauge transformations (\ref{gtr}) are restored with help of
$\delta \ffi^a=\frac{\delta}{\delta c^{\alpha}} \{ \ffi^a,\Omega \}
\eps^{\alpha}$, the conditions (\ref{cond}) and $\frac{\delta S_{eff}}
{\delta \bar b_{\alpha}}=0$. (The later condition must be used before
differentiation over $c^{\alpha}$. Without it there is only the weak
invariance under independent transformations generated by
$G_{\alpha}$ and $\pi_{\alpha}$.)

Having in mind this classical situation one can impose the conditions
(\ref{cond}) on the average fields to extract the gauge invariant
effective action. Define
\be \Gamma=\Gamma_{eff}\vert_{\eta=\bar \eta=0 \atop
 \pi=0 \ \ } \qquad
     \Lambda^{i}_{(0)}=[\Lambda^{i}]_{lg},
     \label{Gam} \ee
where the subscript "lg" means a linear on ghost part. From (\ref{Om})
and (\ref{glsym}) one obtains the explicit expressions:
\beq
\Lambda^{(\eta)}_{(0)}=\langle \{\eta,\Omega\}_+\rangle_{lg}=0,&\ &
\Lambda^{(\bar c_{\alpha})}_{(0)}=\langle \{\bar c_{\alpha},\Omega\}_+
\rangle_{lg}=i\pi_{\alpha},\nonumber \\
\Lambda^{(\pi_{\alpha})}_{(0)}=\langle \{\pi_{\alpha},\Omega\}
\rangle_{lg}=0,&\ &
\Lambda^{(n^{\alpha})}_{(0)}=\langle \{n^{\alpha},\Omega\}\rangle_{lg}=
-ib^{\alpha}.
\label{trans} \eeq
The first equality is obtained due to conservation of the ghost number.
Then the "lg" part of (\ref{Ward}) with the condition $\pi=0$ yields
\be \frac{\delta \Gamma(\ffi)}{\delta \ffi^a}\Lambda^{(\ffi^a)}_{(0)}=
[J^{\alpha}_{(\bar b)}]_{lg}\Lambda^{(\bar b_{\alpha})}_{(0)}
 \label{gia}.\ee

 Evidently, $[J^{\alpha}_{(\bar b)}]_{lg}=-\frac{\delta^2
\Gamma_{eff}}
{ \delta c^{\beta} \delta \bar b_{\alpha}}\vert_{\eta=\bar \eta=0}
c^{\beta}-\frac{\delta^2 \Gamma_{eff}}
{\delta b^{\beta} \delta \bar b_{\alpha}}\vert_{\eta=\bar \eta=0}
b^{\beta}$.
 So the condition $[J^{\alpha}_{(\bar b)}]_{lg}=0$ can be viewed as
a connection of the parameters of local gauge transformations the role
of which is played by ghosts $c^{\alpha}$ and $b^{\alpha}$. From
(\ref{gia}) these transformations are found to be
\beq
\delta^{(1)}_c z^A=\Lambda^{(z^A)}_{(1)\alpha}c^{\alpha},&\ &
\delta^{(1)}_c n^{\alpha}=0, \nonumber \\
\delta^{(2)}_b z^A=\Lambda^{(z^A)}_{(2)\alpha}b^{\alpha}, &
\ & \delta^{(2)}_b n^{\alpha}=-ib^{\alpha},
\label{tr12} \eeq
where
\be \Lambda^{(z^A)}_{(1)\alpha}=\frac{\delta}{\delta c^{\alpha}}
 \langle\{ z^A,\Omega\}\rangle \vert_{\eta=\bar \eta=0 \atop
 \pi=0 \ \ }, \qquad
    \Lambda^{(z^A)}_{(2)\alpha}=\frac{\delta}{\delta b^{\alpha}}
 \langle\{ z^A,\Omega\}\rangle \vert_{\eta=\bar \eta=0 \atop
 \pi=0 \ \ }. \label{l12} \ee
Thus, provided $[J^{\alpha}_{(\bar b)}]_{lg}=0 \Rightarrow
b^{\alpha}=b^{\alpha}(c)$,
$\Gamma(\ffi)$ is the effective action invariant under
the following quantum gauge transformations
\be \delta^{(q)} \ffi^a=
\delta^{(1)}_{\eps}\ffi^a+\delta^{(2)}_{b(\eps)}\ffi^a
=Q^a_{\alpha}(\ffi)\eps^{\alpha} \label{qge}. \ee
The closed expression for $Q^a_{\alpha}$ can be obtained from
(\ref{tr12}), (\ref{l12}), (\ref{Om}) and the expressions for ghost
propagators.

The analogy of the above mentioned weak invariance (i.e. the
invariance under the transformations (\ref{tr12})) comes again
from (\ref{gia}). For existing of this symmetry
$\Lambda^{(\bar b_{\alpha})}_{(0)}$ should vanish.
This can be done by demanding
\be \frac{\delta \Gamma}{\delta n^{\alpha}}+i\frac{\delta \Gamma}
{\delta z^A}\Lambda^{(z^A)}_{(2)\alpha}=0. \label{ws} \ee
The left--hand side of (\ref{ws}) is an analogy of the classical
constraints $G_{\alpha}$ and (\ref{ws}) should be interpreted as
a weak equality. Note that in general case the contribution
of $q$ and $p$ to this expression does not vanish. So in some sense
there is a mixing between the phase space variables $q^s,p_s$
and the Lagrange multipliers $n^{\alpha}$ due to quantum effects.

 Apart from the quantum gauge transformations (\ref{qge})
one can obtain other symmetry transformations of the effective action.
Let $j_a \equiv J_{(\ffi^a)} $ are sources for the physical fields.
Act by the operator
$\hat T^{\beta,\dots}_{\alpha,\dots}=T^{\beta,\dots}_{\alpha,\dots}
(\ffi,\pi)\vert_{(\ffi,\pi)\to (\frac{\hbar\delta}{i\delta j},
\frac{\hbar\delta}{i\delta J_{(\pi)}}) }$
on the identity (\ref{Ward}), take the "lg" part and impose the
conditions $\pi=0$ and $[J^{\alpha}_{(\bar b)}]_{lg}=0$.
Then we come to the following equation
  \be j_a\frac{\delta}{\delta c^{\lambda}}\left(
\hat T^{\beta,\dots}_{\alpha,\dots} \Lambda^{(\ffi^a)}_{(0)}\right)=
i\hbar \frac{\delta}{\delta c^{\lambda}}\left(
\frac{\delta \hat T^{\beta,\dots}_{\alpha,\dots}}{\delta \ffi^a}
\Lambda^{(\ffi^a)}_{(0)}\right)
-i\frac{\delta [J^{\gamma}_{(\bar c)}]_{lg}}{\delta c^{\lambda}}
\hat T^{\beta,\dots}_{\alpha,\dots}\pi_{\gamma}(j).\label{nward}\ee
The demand that the contraction of the right-hand side of
(\ref{nward}) with some matrix $M_{\beta}^{\lambda}$
vanishes gives rise to new symmetry transformations:
\be \delta \ffi^a=M^{\lambda}_{\beta}
\frac{\delta}{\delta c^{\lambda}}\left(
\hat T^{\beta,\dots}_{\alpha,\dots} \Lambda^{(\ffi^a)}_{(0)}\right)
\eps^{\alpha}=\tilde Q_{\alpha}^a(\ffi)\eps^{\alpha} \label{nqge}. \ee

   Note that we cannot reject the last term in (\ref{nward})
because $\pi$ is differentiated and the result is not proportional
to it. (Vanishing of $\pi$ makes sources dependent and cannot be done
before differentiation.)
Having in mind this remark one can clarify the sense of these
additional transformations. If the condition $\pi=0$ is not hardly
used for making the right--hand side of (\ref{nward}) zero
differentiation with respect to $j_a$ gives
$\tilde Q^a_{\alpha}=-j_b\frac{\delta \tilde Q^b_{\alpha}}
{\delta j_a}$.
Thus new transformations are "trivial" ones vanishing on mass shell.
Nevertheless, as we shall see, they play a certain role.

 Finally, let us discuss the algebra of the quantum gauge
transformations.
Here we cannot give any positive result and we are compelled to
restrict ourselves to general description of the situation.
 For derivation of the algebra one should consider
\be Q^b_{\alpha}\frac{\delta}{\delta \ffi^b}Q^a_{\beta}-
   Q^b_{\beta}\frac{\delta}{\delta \ffi^b}Q^a_{\alpha}, \label{com} \ee
where $Q^a_{\alpha}$ is some quantum average. On the other hand
to extract the classical algebra, which is first approximation
for the quantum algebra, it is necessary to write down the expression
(\ref{com}) as one quantum average. It can be done but the
average of the classical commutator will be only one of many
arising terms. So the quantum algebra will be a deformation of
the classical one.
However in contrast with the usual deformed symmetries, where
the algebra is closed in its enveloping algebra, there are additional
"trivial" symmetries which can contribute to (\ref{com}).
Of course, on mass shell they must disappear.
 The situation is similar for the quantum BRST transformation.
There is no reason for it to be nilpotent.

\section{Background gauges}
  Now we introduce in the formalism background fields. It can be done
in two ways. First, one can split the quantum fields $\psi$ into the
classical part $\bF$ and quantum fluctuations. Second, one can take
the gauge fixing function dependent on them. We shall not go by
the first way for the following reason.
It implies that quantum fluctuations are small
in some sense and one can demand their vanishing as an
invariant condition under gauge transformations.
As a result the gauge invariant effective action depending only on
background fields can be obtained. In our case however after
subtraction of the classical part from an average field the result
will not be transformed homogeneously under the quantum gauge
transformations and the onefield construction fails.

 Thus we work with the full average fields $\psi$ and
dependence on the background fields $\bF$ comes only from the function
$\Psi$. The effective action is a functional of variables of two types,
one of them playing a role of external parameters. Since
the above--stated considerations don't depend on presence of such
parameters the effective action is still invariant under (\ref{glsym})
or  (\ref{qge}).

 It is proved that the presence of
background fields gives rise to a new local symmetry including
their transformations. Namely, any variation of the background fields
can be compensated by an appropriate transformation of the average
fields. Indeed, a variation $\delta \bF$ leads to
\be \delta W(J,\bF)=
\frac{i}{\hbar}\langle \{ \delta \Psi(\psi,\bF),\Omega\}_+
\rangle .\ee
On the other hand the BRST transformation of integrated variables
in $\langle \delta \Psi \rangle$ gives
  \be \langle \{ \delta \Psi,\Omega \} \rangle =\frac{i}{\hbar}
  \langle J_i\Omega^i \delta\Psi  \rangle. \ee
Comparing these two equality, it is easy to see that the effective
action  is invariant under the following transformations
\be \delta \bF^i=\eps^i, \qquad \delta \psi^i=\frac{i}{\hbar}
\langle \{ \psi^i,\Omega\}
  \frac{\delta \Psi}{\delta \bF^j}
 \rangle \epsilon^j=\Lambda^{(\bF)i}_j\eps^j.
\label{bqe} \ee

 Now one can return to the physical sector of the theory. For this
nullify all ghosts and $\pi$ in the symmetry equation for
$\Gamma_{eff}$. Then one obtains
    \be \frac{\delta \Gamma(\ffi,\bF)}{\delta \bF^a}+
    \frac{\delta \Gamma(\ffi,\bF)}{\delta \ffi^b}Q^{(\bF)b}_a=0, \ee
where $Q^{(\bF)b}_a=\Lambda^{(\bF)b}_a\vert_{\eta=\bar \eta=0 \atop
\pi=0 \ \ }$ and throughout background fields are
introduced for the physical fields only. This identity means that
the gauge invariant effective action possesses the additional local
symmetry
\be \delta \bF^i=\eps^i, \qquad \delta \ffi^a=Q^{(\bF)a}_b\eps^b.
\label{bge} \ee

  As a result we have the set of local symmetries of $\Gamma$: the
quantum gauge transformations (\ref{qge}), the "trivail" symmetries
(\ref{nqge}) and the transformations induced by background fields
(\ref{bge}).
Since any their linear combination is also a symmetry
transformation, one can try to find such combination which
has some standard form, for example, classical one. It
is clear that it cannot be achieved without use of background fields.
So it is natural to consider the gauge transformations of
the following kind
 \be \delta \bF^a=G^a_{\alpha}(\bF)\eps^{\alpha}, \
 \delta \ffi^a=\left( Q^{(\bF)a}_b G^b_{\alpha}(\bF)+
 Q^a_{\beta}X^{\beta}_{\alpha}(\ffi,\bF)+
\hat Q^a_{\beta}Y^{\beta}_{\alpha}(\ffi,\bF) \right)\eps^{\alpha}
=Q^{(tot)a}_{\alpha}\eps^{\alpha} .
\label{tge} \ee
As we shall see it is possible for the case of the Yang--Mills theory
to find the coefficients
$X^{\beta}_{\alpha}, Y^{\beta}_{\alpha}$ (and the function $\Psi$),
which reduce (\ref{tge}) to the classical gauge transformation,
what reproduces known DeWitt's result in our formalism. For more
general rank one theories the natural generalization also is possible.

\section{Rank one theory}
 let us consider the rank one theory in which the structure functions
$U^{(n)}$ vanish for $n\ge 2$ and the expansion of the BRST
invariant Hamiltonian contains only two terms (\ref{Ham}).
Choose the gauge fixing function in the form (\ref{psi}) with
\be \chi^{\alpha}=\frac{\alpha}{2}(\gamma(\bF)^{-1})^{\alpha \beta}
\pi_{\beta}+\partial_t n^{\alpha}+\xi^{\alpha}(\ffi,\bF).
 \label{gauge}\ee
Then the equation (\ref{Seff}) reads
\beq S_{eff}&=& S(q,p,n)-\frac{\alpha}{2}
(\gamma^{-1})^{\alpha \beta}\pi_{\alpha}\pi_{\beta}-
\pi_{\alpha}\xi^{\alpha}  +\dot c^{\alpha}\bar b_{\alpha}
+ib^{\alpha}\bar b_{\alpha}-b^{\alpha}\frac{\delta \xi^{\beta}}
{\delta n^{\alpha}}\bar c_{\beta}
 \nonumber  \\  &&
-c^{\alpha}V_{\alpha}^{\beta}\bar b_{\beta}
+c^{\alpha}n^{\gamma}C_{\alpha \gamma}^{\beta} \bar b_{\beta}+
ic^{\alpha}\{ G_{\alpha},\xi^{\beta}\}\bar c_{\beta}+
\frac{i}{2}c^{\alpha}c^{\beta}\{ C_{\alpha \beta}^{\gamma},
\xi^{\lambda} \}\bar b_{\gamma}\bar c_{\lambda}  . \eeq
   Integration over $b,\ \bar b$ and $\pi$ in (\ref{zed})  gives
\beq Z(j,J_{\pi},J^{\eta},J_{\eta})&=&
\int {\cal D}\ffi {\cal D}\pi
{\cal D}c{\cal D}\bar c\, \exp\Bigl\{ \frac{i}{\hbar}
\Bigl( S(\ffi)+\frac{1}{2\alpha}\gamma_{\alpha \beta}(\xi^{\alpha}-
J_{(\pi)}^{\alpha})(\xi^{\beta}-J_{(\pi)}^{\beta})+j^a\ffi_a
\Bigr. \Bigr.\nonumber \\
&&
+i\bar c_{\beta}F^{\beta}_{\alpha}c^{\alpha}+(J^{(c)}_{\alpha}+
iJ^{(b)}_{\beta}(\delta^{\beta}_{\alpha}\partial_t+
n^{\gamma}C^{\beta}_{\alpha \gamma}-V_{\alpha}^{\beta}))c^{\alpha}
-\bar c_{\beta}(J^{\beta}_{(\bar c)}-i\frac{\delta \xi^{\beta}}
{\delta n^{\alpha}}J_{(\bar b)}^{\alpha})\nonumber \\
&&+iJ_{\alpha}^{(b)} J_{(\bar b)}^{\alpha}
\Bigl. \Bigl. +\frac{1}{2}J^{(b)}_{\gamma}\bar c_{\lambda}\{
C^{\gamma}_{\alpha \beta},\xi^{\lambda}\} c^{\alpha}c^{\beta}
+\frac{1}{2}\frac{\delta \xi^{\delta}}{\delta n^{\gamma}}
\bar c_{\delta}\bar c_{\lambda}
\{ C^{\gamma}_{\alpha \beta},\xi^{\lambda}\} c^{\alpha}c^{\beta}
     \Bigr) \Bigr\}, \label{basic} \eeq
where
\be F^{\beta}_{\alpha}=\{ \xi^{\beta},G_{\alpha}\}+
\frac{\delta \xi^{\beta}}{\delta n^{\gamma}}
(\delta^{\gamma}_{\alpha}\partial_t+n^{\lambda}C^{\gamma}_{\alpha
\lambda} -V^{\gamma}_{\alpha})=\frac{\delta \xi^{\beta}}{\delta
\ffi^a}G^a_{\alpha}  \label{det} \ee
and we omit factors dependent only on background fields since they
are canceled by normalization.

 Impose the conditions on the gauge:
$ \{C^{\gamma}_{\alpha \beta},
\xi^{\lambda}\}\frac{\delta \xi^{\delta}}{\delta n^{\gamma}}=0$
and $\frac{\delta \xi^{\alpha}}{\delta n^{\beta}}$ does not
depend on $\ffi$. Due to this the term of fourth order in ghosts
in (\ref{basic}) disappears and we can find expressions for
the quantum gauge transformations without ghosts. Note that
from (\ref{basic}) it follows
 \be c^{\alpha}=-i\langle (F^{-1})^{\alpha}_{\beta}
 (J_{(\bar c)}^{\beta}
-i\frac{\delta \xi^{\beta}}{\delta n^{\gamma}}J_{(\bar b)}^{\gamma}
+\frac{1}{2}\{ C^{\lambda}_{\gamma \delta},\xi^{\beta}\}c^{\gamma}
c^{\delta}J^{(b)}_{\lambda} ) \rangle_{lg}=
-i[J_{(\bar c)}^{\beta}-i\frac{\delta \xi^{\beta}}{\delta
n^{\gamma}}J_{(\bar b)}^{\gamma}]_{lg}
\langle (F^{-1})^{\alpha}_{\beta}\rangle_0.   \label{cc}
\ee
 Here we introduced the notation $\langle \cdot \rangle_0\equiv
\langle \cdot \rangle\vert_{\eta=\bar \eta =0}$. With this equality
 one obtains

 \beq  \Lambda^{(z^A)}_{(0)}&=&\langle c^{\alpha}\{z^A,G_{\alpha}\}
 +\frac12c^{\alpha}c^{\beta}\{z^A,C^{\gamma}_{\alpha \beta}\}
 \bar b_{\gamma}\rangle_{lg}
 =-i[J_{(\bar c)}^{\beta}-i\frac{\delta \xi^{\beta}}{\delta
n^{\gamma}}J_{(\bar b)}^{\gamma}]_{lg}
\langle (F^{-1})^{\alpha}_{\beta}\{z^A,G_{\alpha}\} \rangle_0
\nonumber \\
&&-\frac12 \left[
e^{-\frac{i}{\hbar}W}\{z^A,C^{\gamma}_{\alpha \beta}\}
\vert_{\ffi \to \frac{\hbar \delta}{i\delta j}}\frac{\hbar \delta}
{i\delta J_{(\bar b)}^{\gamma}}(F^{-1})^{\alpha}_{\tau}
(J_{(\bar c)}^{\tau}-i\frac{\delta \xi^{\tau}}{\delta
n^{\lambda}}J_{(\bar b)}^{\lambda})
(F^{-1})^{\beta}_{\sigma}
(J_{(\bar c)}^{\sigma}-i\frac{\delta
\xi^{\sigma}}{\delta n^{\delta}}J_{(\bar b)}^{\delta})
e^{\frac{i}{\hbar}W}\right]_{lg} \nonumber \\
&=&  \left(
\langle (F^{-1})^{\gamma}_{\beta}\{z^A,G_{\gamma}\} \rangle_0
-i\hbar
\langle (F^{-1})^{\gamma}_{\beta}\{z^A,C^{\tau}_{\gamma \lambda}\}
(F^{-1})_{\sigma}^{\lambda}\rangle_0
\frac{\delta \xi^{\sigma}}{\delta n^{\tau}} \right)
\left( \langle F^{-1}\rangle_0^{-1}\right)_{\alpha}^{\beta} c^{\alpha}
\label{trq} \eeq
\beq \Lambda_{(0)}^{(n^{\alpha})}&=&-ib^{\alpha}=
\langle (\delta^{\alpha}_{\beta} \partial_t +n^{\gamma}
C^{\alpha}_{\beta \gamma}-V^{\alpha}_{\beta})c^{\beta}\rangle_{lg}
+\left[ J_{(\bar b)}^{\alpha}-
\frac{i}{2} \langle \bar c_{\lambda} \{ C^{\alpha}_{\beta \gamma},
\xi^{\lambda} \} c^{\beta}c^{\gamma} \rangle \right]_{lg}
\nonumber \\
& \mathop{=}\limits_{[J_{(\bar b)}^{\alpha}]_{lg}=0}
& \left(
\langle (F^{-1})^{\gamma}_{\beta}(\delta^{\alpha}_{\gamma}\partial_t+
n^{\lambda}C^{\alpha}_{\gamma \lambda}-V^{\alpha}_{\gamma}) \rangle_0
-i\hbar\langle (F^{-1})^{\gamma}_{\beta}\{C^{\alpha}_{\gamma \lambda}
,\xi^{\sigma} \} (F^{-1})_{\sigma}^{\lambda}\rangle_0 \right)
\left( \langle F^{-1}\rangle_0^{-1}\right)_{\tau}^{\beta} c^{\tau}
\label{trn} \eeq
Since the condition $[J_{(\bar b)}^{\alpha}]_{lg}=0$ was not used
in (\ref{trq}) we have $\Lambda^{(z^A)}_{(2)\alpha}=0$.
So in this case there is no a mixing between the Lagrange multipliers
and the phase space variables,
and the weak invariance of the effective
action is guaranteed by the standard equation
$\frac{\delta \Gamma}{\delta n^{\alpha}}=0$.
This is a direct consequence of the above imposed
conditions on the gauge fixing function.

  The transformations (\ref{trq}), (\ref{trn}) can be incorporated
into one expression:
\beq Q^a_{\alpha}&=&
\left(
\langle (F^{-1})^{\gamma}_{\beta}G^a_{\gamma} \rangle_0
-i\hbar
\langle (F^{-1})^{\gamma}_{\beta}\{\ffi^a,C^{\tau}_{\gamma \lambda}\}
(F^{-1})_{\sigma}^{\lambda}\rangle_0
\frac{\delta \xi^{\sigma}}{\delta n^{\tau}}
\right. \nonumber \\
&&\left. +i\hbar\langle (F^{-1})^{\gamma}_{\beta}
\{ \xi^{\sigma},\frac{\delta G^a_{\gamma}}{\delta n^{\lambda}} \}
(F^{-1})_{\sigma}^{\lambda}\rangle_0  \right)
\left( \langle F^{-1}\rangle_0^{-1}\right)_{\alpha}^{\beta}
 \nonumber \\
&=& \left(
\langle (F^{-1})^{\gamma}_{\beta}G^a_{\gamma} \rangle_0
-i\hbar
\langle (F^{-1})^{\gamma}_{\beta}
\frac{\delta }{\delta n^{\lambda}} \left(
\{\ffi^a,F^{\sigma}_{\gamma}\} - \{ \xi^{\sigma},G^a_{\gamma} \}
\right)
(F^{-1})_{\sigma}^{\lambda}\rangle_0 \right)
\left( \langle F^{-1}\rangle_0^{-1}\right)_{\alpha}^{\beta}.
\label{qgtr} \eeq

 To obtain the "trivial" gauge transformations let us take
 \be T_{\alpha}^{\beta}(\ffi,\pi)=-\frac{\alpha}{2}
 \left( C^{\beta}_{\alpha \gamma}(\bF)(\gamma^{-1})^{\gamma \lambda}-
 (\gamma^{-1})^{\beta \gamma}C^{\lambda}_{\alpha \gamma}(\bF)\right)
 \pi_{\lambda}, \qquad
 M^{\lambda}_{\beta}=\langle (F^{-1})^{\lambda}_{\beta}\rangle_0. \ee
It is easy to see that due to (\ref{cc})
the contraction of the right--hand side of (\ref{nward}) with $M$
vanishes. Thus the "trivial" gauge transformations are given by
\beq \tilde Q^a_{\alpha}&=&\frac12
\left( \langle (F^{-1})^{\gamma}_{\beta}G^a_{\gamma}(\xi^{\rho}
-J_{(\pi)}^{\rho}) \rangle_0
-i\hbar \langle (F^{-1})^{\gamma}_{\beta}
\frac{\delta }{\delta n^{\lambda}} \left(
\{\ffi^a,F^{\sigma}_{\gamma}\} - \{ \xi^{\sigma},G^a_{\gamma} \}
\right) (F^{-1})_{\sigma}^{\lambda}
(\xi^{\rho}-J_{(\pi)}^{\rho})\rangle_0
\right)
\nonumber \\ && \times
\left( C^{\beta}_{\alpha \rho}(\bF)-
 (\gamma^{-1})^{\beta \tau}C^{\delta}_{\alpha \tau}(\bF)
 \gamma_{\delta \rho}\right)
 \nonumber \\
&=&\frac12
\left( \langle (F^{-1})^{\gamma}_{\beta}G^a_{\gamma}\xi^{\rho}
\rangle_0
-i\hbar \langle (F^{-1})^{\gamma}_{\beta}
\frac{\delta }{\delta n^{\lambda}} \left(
\{\ffi^a,F^{\sigma}_{\gamma}\} - \{ \xi^{\sigma},G^a_{\gamma} \}
\right) (F^{-1})_{\sigma}^{\lambda}\xi^{\rho}\rangle_0
\right)
\nonumber \\ && \times
\left( C^{\beta}_{\alpha \rho}-
 (\gamma^{-1})^{\beta \tau}C^{\delta}_{\alpha \tau}
 \gamma_{\delta \rho}\right)
 -\frac12 Q^a_{\gamma}\langle (F^{-1})^{\gamma}_{\beta}\rangle_0
\left( C^{\beta}_{\alpha \rho}-
 (\gamma^{-1})^{\beta \tau}C^{\delta}_{\alpha \tau}
 \gamma_{\delta \rho}\right) \langle \xi^{\rho}\rangle_0.
\label{triv} \eeq
In the last equality the condition $\pi_{\alpha}=\frac{1}{\alpha}
\gamma_{\alpha \beta}
(J_{(\pi)}^{\beta}-\langle \xi^{\beta} \rangle)=0$ was used.

 The transformations induced by background fields are found from
(\ref{bqe}) to be
\beq Q^{(\bF)a}_b&=&\frac{1}{\hbar}\left\langle
\{ \ffi^a,\Omega\}\bar c_{\beta}
\left(\frac{\alpha}{2}(\gamma^{-1})^{\beta \tau}\frac{\delta
\gamma_{\tau \sigma}}{\delta \bF^b}(\gamma^{-1})^{\sigma \lambda}
\pi_{\lambda}-\frac{\delta \xi^{\beta}}{\delta \bF^b}\right)
\right\rangle_0
\nonumber \\
&=& -
\left\langle (F^{-1})^{\gamma}_{\beta}G^a_{\gamma}
\left( \frac12 (\gamma^{-1})^{\beta \tau}\frac{\delta
\gamma_{\tau \lambda}}{\delta \bF^b}(\xi^{\lambda}-J_{(\pi)}^{\lambda})
+\frac{\delta \xi^{\beta}}{\delta \bF^b}\right)
\right\rangle_0
 \nonumber \\ &&
+i\hbar \left\langle (F^{-1})^{\gamma}_{\beta}
\frac{\delta }{\delta n^{\lambda}} \left(
\{\ffi^a,F^{\sigma}_{\gamma}\} - \{ \xi^{\sigma},G^a_{\gamma} \}
\right) (F^{-1})_{\sigma}^{\lambda}
\left( \frac12 (\gamma^{-1})^{\beta \tau}\frac{\delta
\gamma_{\tau \rho}}{\delta \bF^b}(\xi^{\rho}-J_{(\pi)}^{\rho})
+\frac{\delta \xi^{\beta}}{\delta \bF^b}\right)
\right\rangle_0 \label{bactr}
\eeq

 Let the following covariance conditions are satisfied:
\be \frac{\delta \gamma_{\alpha \beta}}{\delta \bF^a}
G^a_{\gamma}(\bF)=-\gamma_{\beta \lambda}C^{\lambda}_{\gamma \alpha}
(\bF)-\gamma_{\alpha \lambda}C^{\lambda}_{\gamma \beta}(\bF), \ee
\be  \frac{\delta \xi^{\alpha}}{\delta \ffi^a}G^a_{\beta}(\ffi)+
\frac{\delta \xi^{\alpha}}{\delta \bF^a}G^a_{\beta}(\bF)=
C^{\alpha}_{\beta \gamma}(\bF)\xi^{\gamma} \label{covar}. \ee
 The later is a generalization of the well known condition for the
gauge fixing function for the Yang--Mills theory. It does not look
very natural, but just it gives the more or less natural
expression for the symmetry transformation of the effective action.
>From (\ref{bactr}), (\ref{qgtr}) and (\ref{triv})  we have
 \beq  Q^{(\bF)a}_bG^b_{\alpha}(\bF)&=&
\left\langle \left( (F^{-1})^{\gamma}_{\beta}G^a_{\gamma}
-i\hbar (F^{-1})^{\gamma}_{\beta}
\frac{\delta }{\delta n^{\lambda}} \left(
\{\ffi^a,F^{\sigma}_{\gamma}\} - \{ \xi^{\sigma},G^a_{\gamma} \}
\right) (F^{-1})_{\sigma}^{\lambda} \right)
\right. \nonumber \\ && \left. \times
\left(
\frac{\delta \xi^{\beta}}{\delta \ffi^b}G^b_{\alpha}-
\frac12 \left( C^{\beta}_{\alpha \rho}(\bF) -
(\gamma^{-1})^{\beta \tau}C^{\delta}_{\alpha \tau}(\bF)
\gamma_{\delta \rho} \right)\xi^{\rho}
+\frac12 (\gamma^{-1})^{\beta \tau}\frac{\delta
\gamma_{\tau \rho}}{\delta \bF^b}\langle\xi^{\rho}\rangle_0
\right) \right\rangle_0
\nonumber \\
&=& \langle G^a_{\alpha}\rangle_0
-i\hbar \langle \frac{\delta }{\delta n^{\lambda}} \left(
\{\ffi^a,F^{\beta}_{\alpha}\} - \{ \xi^{\beta},G^a_{\alpha} \}
\right) (F^{-1})_{\beta}^{\lambda} \rangle_0
-\tilde Q^a_{\alpha}
-Q^a_{\beta}\langle (F^{-1})^{\beta}_{\gamma}\rangle_0
C^{\gamma}_{\alpha \lambda}  \langle \xi^{\lambda}\rangle_0
\label{bgtr}.  \eeq
Thus if we shall choose in (\ref{tge}) $X^{\beta}_{\alpha}=
\langle (F^{-1})^{\beta}_{\gamma}\rangle_0
C^{\gamma}_{\alpha \lambda}  \langle \xi^{\lambda}\rangle_0$
and $Y^{\beta}_{\alpha}=\delta^{\beta}_{\alpha}$ then the total
quantum gauge transformation reads
\be Q^{(tot)a}_{\alpha}=\langle G^a_{\alpha}\rangle_0
-i\hbar \langle \frac{\delta }{\delta n^{\lambda}} \left(
\{\ffi^a,F^{\beta}_{\alpha}\} - \{ \xi^{\beta},G^a_{\alpha} \}
\right) (F^{-1})_{\beta}^{\lambda} \rangle_0. \label{total} \ee

 In the case of the  Yang--Mills theory the second term is absent, and
since $G_{\alpha}^a$ is linear in the fields the quantum
transformation is reduced to the classical one:
$Q^{(tot)a}_{\alpha}=G^a_{\alpha}$, i.e. $\Gamma(\ffi,\bF)$ is
the classically gauge invariant effective action.
In more general case the natural generalization is
$Q^{(tot)a}_{\alpha}=\langle G^a_{\alpha}\rangle_0$, that
corresponds to the case of a global symmetry (\ref{glsym}).
This result can be achieved by vanishing of the second term.
One needs two additional conditions on the gauge fixing function:
$\frac{\delta \xi^{\alpha}}{\delta n^{\beta}}=0$ and
$\{ \xi^{\alpha},C^{\lambda}_{\beta \gamma}\}=0$.
They are very strong if the structure constants depend on all
coordinates and momenta. As example of such theory
the Ashtekar gravity can be pointed out \cite{ash}.
On the other hand in the ADM gravity \cite{adm}
the conditions forbid gauges on momenta only.

 It may seem that the second term in (\ref{total}) vanishes due to
a renormalization procedure. If all operators are local the contraction
of $\lambda$ and $\beta$ would result in appearance of $\delta(0)$,
which in the dimensional regularization should be put zero. However
the presence of the ghost propagator $(F^{-1})^{\lambda}_{\beta}$
can give rise to nonlocality and the reasoning fails.

 Finally we consider the problem of $\delta$--gauges, i.e. gauges
leading to $\delta$--functions. They are obtained from
(\ref{gauge}) in the limit $\alpha \longrightarrow 0$.
It is easy to see that
in this limit the effective action is also invariant
under the transformations (\ref{total}) under the same conditions
on the gauge fixing function $\xi^{\alpha}$. However the question
can arise: does the propagator $\frac{\delta^2 W}
{\delta J \delta J}$ remain non-degenerate?
Note that the argument of $\delta$--function is always
nonhomogeneous due to appearance of $J_{(\pi)}$.
Its presence is meaningful since just it guarantees
non-degeneracy. So the limit to $\delta$--gauge  is well defined.

 It opens new possibilities for simplification of the quantum
gauge transformations. Using $\delta$--gauges one can make
the ghost propagator $F_{\alpha}^{\beta}$ independent on fields.
Then provided the second term in (\ref{qgtr}) vanishes we come to
$Q^a_{\alpha}=\langle G^a_{\alpha}\rangle_0$, that gives the classical
symmetry for the Yang--Mills theory without use of background fields.
 In this case the result can be
achieved with help of such noncovariant gauges as $\xi=A_3$ or $\xi=A_0$
(here $A_{\mu}$ is the gauge potential).

\section{Conclusions and Discussion}

  In this article we have considered symmetry properties of
the effective action for general gauge theories. By taking as a basis
the Hamiltonian BRST formalism and the path integral quantization
we have found the quantum gauge transformations in terms of
quantum averages. Moreover there are additional symmetry
transformations of the effective action: the "trivial" symmetries,
which vanish on mass shell, and the gauge transformations induced by
background fields. (We don't identify background fields with
average fields so that the effective action is a functional of these
two variables.) Finally we have shown that combining all these
symmetries one can obtain more simple form of the transformations.
For example, for the rank one theory under some conditions on
the gauge fixing function the gauge transformations are represented
as an average of the classical ones. For Yang--Mills theory, in which
the gauge transformations are linear in the fields, it gives
the classical result for the quantum symmetry also.

 Unfortunately this generalization for the rank one theory is
very weak because of the covariance conditions (\ref{covar}).
Their solution is a problem and, may be, the result isn't worth the
efforts. As we have seen this is only the attempt to find some simple
form of the symmetry, so there is no need to introduce new
difficulties. On the other hand this formalism translates
the problem of search for the (classically) invariant effective action
to the problem of solving of equations.

 We did not concern of the renormalization problem. It becomes
complicated due to that composite operators enter into expressions
for the quantum gauge transformations. However this is the standard
difficulty for investigations which deal with the Ward identities.
So we suppose that in real calculations all expressions should be
renormilized taking into account a mixing of operators.

 Of course, such calculations should be carried out with help of
expansion in $\hbar$. Each order together with previous ones gives
restrictions to the corresponding order of expansion of the effective
action. It is worth to notice that the first  nontrivial correction
to the classical transformations is fully defined by the classical
action since it is proportional
$\sim \hbar \frac{\delta^2 W}{\delta J \delta J} \sim
\hbar \left( \frac{\delta^2 \Gamma}{\delta \psi \delta \psi}\right)^{-1}
\sim \hbar\left( \frac{\delta^2 S}{\delta \psi \delta \psi}\right)^{-1}$.

 The consideration of this article can be translated to the Lagrangian
BV formalism. The proof of its equivalence to the Hamiltonian BFV
formalism can be found for instance in \cite{equiv}.
Within this framework all symmetry properties
of the effective action are contained in the Zinn--Justin equation
\cite{zj} $(\Gamma,\Gamma)=0$,
where $(\, ,\, )$ is the antibracket and
$\Gamma$ is a functional of fields and antifields
\cite{bv,bigrev,htbook}. Since it has the same form as the
classical master equation $(S,S)=0$ one can define a BRST structure
associated with $\Gamma$ \cite{anom}.
However there are small difficulties similar to difficulties in
the Hamiltonian approach. Whereas for the proper solution $S$
ghost number considerations allow to conclude that
$S_0$ is the local invariant classical action,
we cannot maintain this for $\Gamma_0$. (Here the subscript 0 means
vanishing all ghosts and antifields.)
This is bacause in the quantum case to fix the gauge we must
introduce auxilary fields. The antifield ${\bar C}^{\alpha *}$
for one of them has ghost number zero \cite{bigrev}.
So $\Gamma_0$ will be local invariant only on the surface
$\left( \frac{\partial \Gamma}{\partial {\bar C}^{\alpha *}}
\right)_0=0$.
It should be viewed as an equation on the auxilary
field ${\bar \pi}_{\alpha}$ which plays a role of momentum
conjugated to Lagrange multiplier
and has ghost number zero.
This corresponds to necessity of removal all auxilary
fields and gives the concret way for this.
Our choice of the Hamiltonian formalism is connected to wish to avoid
the problem of solution of the quantum master equation
\cite{bv,bigrev,htbook}. However notice that
in BV approach the quantum algebra may be simpler
then in our case \cite{lv,qtrans}.

 Another progress can be connected with an application of
quantum groups. It is just the structure that should manage
the quantum symmetry that is a deformation of the classical one.

\section*{Acknowledgments}

 The author is very grateful to D.V.Vassilevich and V.A.Franke for
helpful and valuable discussions.


\begin{references}

\bibitem{renom}
't Hooft G., Nucl. Phys. {\bf B33}, 173 (1971), {\bf B35}, 167 (1971).
\bibitem{dewitt}
 B.S.DeWitt, in: {\it Quantum Gravity 2}, eds. C.J. Isham,
 R.Penrose and D.W.Sciama. (Clarendon Press, Oxford, 1981).
\bibitem{all}
L.F.Abbott, Nucl. Phys. {\bf B185}, 189 (1981);\\
D.G.Boulware, Phys. Rev. {\bf D 23}, 389 (1981);\\
C.F.Hart, Phys. Rev. {\bf D 28}, 1993 (1983).

\bibitem{vilk2}
G.A.\ Vilkovisky, in:  {\it Quantum Theory of Gravity},
ed. S.M.\ Christensen (Hilger, Bristol 1984).
\bibitem{vilk3}
G.A.\ Vilkovisky, Nucl. Phys. {\bf B234}, 127 (1984).
\bibitem{dewi}
B.S.DeWitt, in: {\it Quantum Field Theory and Quantum
Statistics --- Essays in Honour of the Sixtieth Birthday of
E.S.Fradkin}, eds. I.A.Batalin, C.J.Isham and G.A.Vilkovisky,
(Hilger, Bristol, 1987).

\bibitem{lavr}
 P.M.Lavrov, Mod. Phys. Let. {\bf 6A}, 2051 (1991);\\
 P.M.Lavrov, Phys. Lett. {\bf B 366}, 160 (1996),
 hep-th/9507105 preprint.

\bibitem{lv}
 S.Falkenberg, B.Geyer, P.Lavrov and P.Moshin, Int. J. Mod. Phys.
 {\bf A13}, 4077 (1998), hep-th/9710050 preprint.

\bibitem{BRST}
 C.Becchi, A.Rouet and R.Stora,
Comm. Math. Phys. {\bf 42}, 127 (1975);
Ann. Phys. {\bf 98}, 287 (1976);    \\
 I.V.Tyutin, Lebedev preprint FIAN No.39 (1975).

\bibitem{BFV}
E.S.~Fradkin and G.A.~Vilkovisky, Phys. Lett.
                 {\bf B 55}, 244 (1975);\\
I.A.~Batalin and G.A.~Vilkovisky, Phys. Lett. {\bf B 69}, 309
                 (1977);\\
E.S.~Fradkin and T.E.~Fradkina, Phys. Lett. {\bf B 72}, 343 (1978).

\bibitem{bv}
I.A.Batalin and G.A.Vilkovisky,  Phys. Lett. {\bf B 102}, 27 (1981);
 Phys. Rev. {\bf D 28} 2567 (1983).

\bibitem{vlt}
B.L.Voronov, P.M.Lavrov and I.V.Tyutin,
Sov. J. Nucl. Phys. {\bf 36}, 292 (1982).

\bibitem{qtrans}
A.Sen, B.Zwiebach, Phys. Lett. {\bf B 320}, 29 (1994),
hep-th/9309027 preprint;\\
M.A.Grigoriev, A.M.Semikhatov and I.Yu.Tipunin, hep-th/9804156 preprint.

\bibitem{bigrev}
J.Gomis, J.Paris and S.Samuel, Phys. Rep. {\bf 259}, 1 (1995),
hep-th/9412228 preprint.

\bibitem{htbook}
M.Henneaux and C.Teitelboim,
{\it Quantization of Gauge Systems}
(Princeton University Press, Princeton, 1992).

\bibitem{book}
 B.S.DeWitt, {\it Dynamical Theory of Groups and Fields} (Gordon and
 Breach, New York, 1965).

\bibitem{Henneaux}
M.~Henneaux, Phys. Rep. {\bf 126}, 1 (1985).

\bibitem{anselm}
D.Anselmi, Class. Quant. Grav. {\bf 11}, 2481 (1994),
 hep-th/9309085 preprint.


\bibitem{ash}
A.Ashtekar, Phys. Rev. Lett. {\bf 57}, 2244 (1986);
Phys. Rev. {\bf D 36}, 1587 (1987).
\bibitem{adm}
R.Arnowitt, S.Deser and C.W.Misner, in: {\it Gravitation:
An Introduction to Current Research}, ed. L.Witten, (Wiley, New  York,
1962).

\bibitem{equiv}
C.Batlle, J.Gomis, J.Paris and J.Roca, Nucl. Phys. {\bf B329},
139 (1990);\\
A.Dresse, J.M.L.Fisch, P.Gregoire and M.Henneaux, Nucl. Phys.
{\bf B354} 191 (1991);\\
G.V.Grigorian, R.P.Grigorian and I.V.Tyutin, Sov. J. Nucl. Phys.
{\bf 53} 1058 (1991); Nucl. Phys. {\bf B379}, 304 (1992).


\bibitem{zj}
J.Zinn-Justin,
{\it Renormalization of Gauge Theories},
in Trends in Elementary Particle Theory,
edited by H.Rollnik and K.Dietz,
Lecture Notes in Physics, Vol {\bf 37},
(Springer-Verlag, Berlin, 1975).

\bibitem{anom}
J.Gomis and J.Paris,
Nucl. Phys. {\bf B431}, 378 (1994),
hep-th/9401161 preprint.


\end{references}
\end{document}